# Superradiant Syntheses via the V-type Three-Level Atoms

Gombojav O. Ariunbold [1, *], and Tuguldur Begzjav [2]

[1] Department of Physics and Astronomy, Mississippi State University, 39762 Mississippi State, MS, USA; ag2372@msstate.edu
[2] Department of Physics, National University of Mongolia, 14200 Ulaanbaatar, Mongolia
* Correspondence: ag2372@msstate.edu

**Highlights**

- Exact solutions for two-mode superradiance
- Processing data against the Dicke superradiance characteristics
- Demonstrating a superradiant synthesis
- Potential applications for quantum state engineering

**Abstract**

Quantum state engineering operating with photons is a key enabler of major scientific breakthroughs and future quantum technologies. Its primary obstacle, however, is decoherence often caused by spontaneous emission, which is inherently difficult to control. In contrast, superradiance—a collective burst of radiation from an ensemble of quantum emitters—offers a more controllable alternative. The temporal dynamics of superradiance can be tuned via external experimental parameters, unlike those of spontaneous emission. Incorporating superradiance into quantum state engineering could therefore provide more accessible control over quantum systems. Motivated by this, we present a theoretical model for an ensemble of atoms in a V-type three-level configuration. Our numerical data are analyzed against key superradiance characteristics. The system successively emits a pair of superradiance pulses, effectively synthesizing the dynamically decoupled sub-ensembles into a single macroscopic ensemble involving all atoms. To our knowledge, this process, which we term superradiant synthesis, is demonstrated here for the first time. These findings offer new insights for practical quantum state engineering, particularly in enabling syntheses of macroscopic quantum sub-systems.

## 1. Introduction

Producing, analyzing, controlling and synthesizing quantum states of light to achieve desired target states lies at the core of quantum state engineering. These target states often involve entanglement, made possible by advances in laser technologies, ultracold atoms and superconducting circuits – platforms that are crucial for quantum computing and sensing [1-5]. Recent quantum state engineering methods include drive-and-dissipation, adiabatic cooling, and measurement-based steering [5-12]. However, quantum states are highly susceptible to environmental disturbances, particularly decoherence. A primary source of decoherence is spontaneous emission, a process governed by fundamental quantities that are inherently difficult to control. Spontaneous emission is well-described by the interaction of vacuum field with individual excited two-level atoms [13, 14]. Integrating superradiance into quantum state engineering offers a more diverse and empirically accessible ways to control the temporal dynamics of quantum systems. In this spirit, the preparation of various target quantum states--including Dicke states [15-17]--using the basic principles of superradiance [18-28], has already been proposed, representing as a preliminary study toward advanced quantum state engineering.

In 1954, Robert Dicke theoretically predicted a burst of radiation – superradiance—emanating from a collective atom ensemble [18]. This phenomenon is fundamentally different from the spontaneous emission of individual atoms. In the standard Dicke model of superradiance [23-28], an ensemble of the two-level atoms evolves into a genuine quantum state characterized by a net zero dipole moment [14]. This model has been experimentally realized in systems such as single diamond nanocrystals at room temperature [29]. Unlike spontaneous emission, superradiance provides a more controllable mechanism for light emission. When all $N$ two-level atoms are initially excited, the peak intensity of the superradiant burst scales as $I_{max} \propto N^2$, rather than $N$ [23-30]. The timing of this burst is characterized by a delay $\tau_D$, which decreases with increasing atom number according to the formula: $\tau_D = (E_0 + \log N)/N$, where $E_0 = 0.57721$ is Euler's constant [23-30]. Conversely, $\tau_W$, the temporal width of the burst scales as $\tau_W \propto 1/N$ [19]. Furthermore, $\sigma$, the noise fluctuations of superradiance in the time domain are given by $\sigma = \pi/[\sqrt{6}(E_0 + \log N)]$ [23-28, 30, 31-39].

A natural generalization of Dicke's original model is its extension from a two-level to a three-level atomic system, which permits two-mode superradiance [30, 40-46]. In our recent work [30, 44], we reported exact numerical solutions and detailed analyses of these simplified models from the 1970s [40-43]. When all atoms are initially excited to the uppermost state, this work demonstrated the generation of a timed pair of Dicke superradiance pulses: a cascade superradiance from a ladder configuration [30, 45, 46]. Furthermore, we established a method for the mode-selective control of Dicke superradiance emitted from the Λ-type system [44].

In this work, we present the quantum statistical theory of two-mode superradiance for the remaining three-level configuration: the V-type (Vee) system. We develop a theoretical model for an ensemble of $2N$ atoms arranged in a V-type configuration, initialized with $N$ atoms in one excited state and the remaining $N$ atoms in the other excited state, while none occupy the common ground state. The resulting numerical data are analyzed against the principal characteristics of Dicke superradiance. When the two spontaneous decay rates are chosen to differ by an order of magnitude, the system emits a well-timed, successive pair of superradiance pulses. This behavior effectively synthesizes two dynamically decoupled sub-ensembles of $N$ atoms into a single macroscopic ensemble of $2N$ atoms. To our knowledge, this process--which we term superradiant synthesis-- is demonstrated here for the first time.

Our findings offer new insights for practical quantum state engineering, particularly in enabling syntheses of macroscopic quantum sub-systems. The paper is organized as follows: Section 2 introduces the V-type three-level model and its governing equations. Sections 3 and 4 formulate the observable signals and present the results of numerical simulations, respectively. Section 5 analyzes these results by comparing them with the established properties of Dicke superradiance to provide clarity and interpretation. Finally, Section 6 concludes by discussing the dynamics and implications of this superradiant synthesis.

## 2. Two-mode rate equations

As in our recent works on three-level systems in the ladder [30] and Λ-type configurations [44], we consider an ensemble of V-type three-level atoms confined to a volume smaller than the wavelengths of both radiation modes. Consistent with [23-28, 30, 40-44], dipole-dipole interactions between atoms and the effect associated with the collective frequency shifts are neglected. The atomic configuration consists of two distinct upper (excited) states and a shared lower (ground) state. In the V-type system, one-photon transitions are permitted between the common ground state and each excited state but are forbidden between two excited states. The system dynamics are therefore given by the following rate equations [40-43]:

$$\frac{dP_{n,m}(t)}{dt} = \Gamma_1[I_1(n+1,m+1)P_{n+1,m+1}(t) - I_1(n,m)P_{n,m}(t)] \\ + \Gamma_2[I_2(n,m+1)P_{n,m+1}(t) - I_2(n,m)P_{n,m}(t)]$$

(1)

here $P_{n,m}(t)$ represents the probability of finding the system with $2N$ atoms in the state $|n, m-n, 2N-m\rangle$ at time $t$. The cooperative decay rates for the two distinct one-photon transitions are given by [40-43] : $I_1(n,m) = n(2N-m+1)$ and $I_2(n,m) = (m-n)(2N-m+1)$. In this notation, $2N-m$ denotes the atoms in the shared lower state $|3\rangle$, while $n$ and $m-n$ represent the atoms in the two upper states $|1\rangle$ and $|2\rangle$, respectively (see the inset in Fig. 1). The corresponding single-atom spontaneous decay rates are $\Gamma_1$ and $\Gamma_2$ for the transitions $\lambda_1$ and $\lambda_2$, respectively, with $\lambda_2 \geq \lambda_1$. The initial condition for the entire V-type system of $2N$ atoms is set such that $N$ atoms are excited to state $|1\rangle$, the remaining $N$ atoms are excited to state $|2\rangle$, and none occupy the shared ground state $|3\rangle$.

## 3. Time-dependent two-mode observables

We analyze three time-dependent two-mode observables for the two-mode system: (i) the emission intensities, (ii) the buildup time delays, and (iii) the noise fluctuations. Explicitly, the time-dependent intensities for the two modes are given by [30, 44]:

$$I_{1,2}(t) = \sum_{n,m=0}^{2N} I_{1,2}(n,m) P_{n,m}(t) \tag{2}$$

here, unnecessary constant factors of the transition wavelengths $\lambda_{1,2}$ are omitted [43], as they do not affect the results presented in this work. The time-dependent delay times for the two-mode are given by [30, 44]:

$$\langle \tau_{1,2}(t) \rangle = \frac{1}{\langle A_{1,2}(t) \rangle} \sum_{n,m=0}^{2N} I_{1,2}(n,m) \int_0^t t' P_{n,m}(t') dt' \tag{3}$$

here, the partial time-dependent pulse areas for the two modes are defined as [30, 44]:

$$\langle A_{1,2}(t) \rangle = \sum_{n,m=0}^{2N} I_{1,2}(n,m) \int_0^t P_{n,m}(t') dt' \tag{4}$$

where the corresponding total pulse areas for the two modes are given by:

$$\langle A_{1,2}(\infty) \rangle = \sum_{n,m=0}^{2N} I_{1,2}(n,m) \int_0^\infty P_{n,m}(t') dt' \tag{5}$$

In the long-time limit ($t \to \infty$), these converge to

$$\langle \tau_{1,2}(\infty) \rangle = \frac{1}{\langle A_{1,2}(\infty) \rangle} \sum_{n,m=0}^{2N} I_{1,2}(n,m) \int_0^\infty t' P_{n,m}(t') dt' \tag{6}$$

where the normalization uses the total pulse areas. Finally, the time-dependent two-mode noise fluctuations in time domain are quantified by [30, 44]

$$\langle \sigma_{1,2}(t) \rangle = \frac{\sqrt{\langle \tau_{1,2}(t)^2 \rangle - \langle \tau_{1,2}(t) \rangle^2}}{\langle \tau_{1,2}(t) \rangle} \tag{7}$$

where the second moment is calculated as [30, 44]

$$\langle \tau_{1,2}(t)^2 \rangle = \frac{1}{\langle A_{1,2}(t) \rangle} \sum_{n,m=0}^{2N} I_{1,2}(n,m) \int_0^t t'^2 P_{n,m}(t') dt' \tag{8}$$

The intrinsic noise fluctuations in the system originate from quantum interactions between atoms and vacuum fluctuations [31, 32]. Numerical results obtained from Eqs. (1-8) are presented in the next section.

## 4. Numerical simulations

In this section, we numerically compute the time-dependent intensities, buildup time delays, and noise fluctuations for the two-modes. Using Eq. (2) and the solutions for $P_{n,m}(t)$ from Eq. (1), the two-mode intensities are calculated as functions of the number of atoms $N$ and time $t$, with the results depicted in Fig. 1. The figure shows that both modes exhibit pulsed behavior. All emitted pulses are delayed, emerging with a rapidly decreasing buildup time as the number of atoms $N$ increases. The detailed temporal characteristics of these pulsed modes are discussed in the following section.

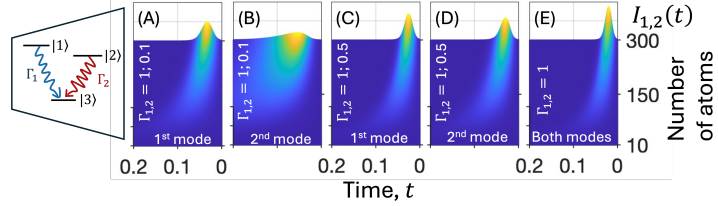

**Figure 1.** Temporal evolution of the two-mode intensities $I_{1,2}(t)$ as functions of the number of atoms. Results are shown for different decay rates: (A, B) $\Gamma_1 = 1$ and $\Gamma_2 = 0.1$; (C, D) $\Gamma_1 = 1$ and $\Gamma_2 = 0.5$; and (E) $\Gamma_1 = \Gamma_2 = 1$. The total number of three-level atoms is $2N$, with $N$ varying discreetly as integers in the range $[5, 150]$. Panels (A) and (C) correspond to the first radiation mode, while (B) and (D) show the intensities for the second mode. Panel (E) is identical for both modes. The inset is schematics for V-three-level configuration, where $n$ atoms are in the upper state $|1\rangle$, $m - n$ atoms are in the upper state $|2\rangle$, and $2N - m$ atoms in are in the shared ground state $|3\rangle$.

Figure 2 shows the numerical results for the temporal evolution of the two-mode buildup time delays $\langle \tau_{1,2}(t) \rangle$ as functions of the number of atoms $N$. The delay profiles exhibit qualitatively similar behavior for both modes. As mentioned in [44], the delays first decrease sharply with increasing atom number and then converge, transitioning to a steady-state value at long times ($t \to \infty$). This temporal behavior was examined in detail in [44].

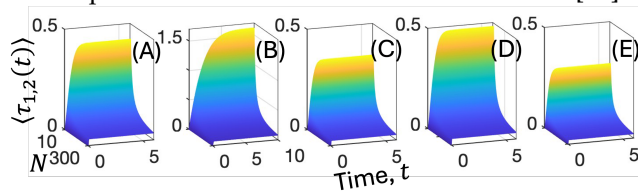

**Figure 2.** Temporal evolution of the two-mode time delays $\langle \tau_{1,2}(t) \rangle$ as functions of the number of atoms. Results are shown for different decay rates: (A, B) $\Gamma_1 = 1$ and $\Gamma_2 = 0.1$; (C, D) $\Gamma_1 = 1$ and $\Gamma_2 = 0.5$; and (E) $\Gamma_1 = \Gamma_2 = 1$. The total number of three-level atoms is $2N$, with $N$ varying discretely as integers $N \in [5, 150]$. Panels (A) and (C) correspond to the first radiation mode, while (B) and (D) show the time delays for the second mode. Panel (E) is identical for both modes.

Figure 3 presents the time-dependent two-mode noise fluctuations $\langle \sigma_{1,2}(t) \rangle$ as functions of $N$. Similarly to the behavior observed in Refs.[30, 44], the noise fluctuations exhibit the following temporal behavior. First, they abruptly decrease, exhibiting "valleys", and then become stationary at longer times. The origins of these valleys and the asymptotic behavior in the limit $t \to \infty$ were discussed in Ref. [44].

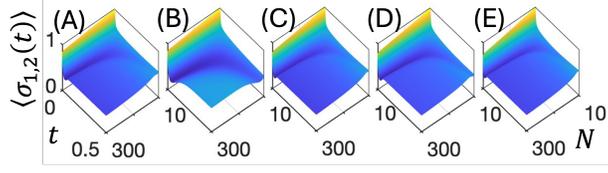

**Figure 3.** Temporal evolution of the noise fluctuations $\langle \sigma_{1,2}(t) \rangle$ as functions of the number of atoms. Results are shown for different decay rates: (A, B) $\Gamma_1 = 1$ and $\Gamma_2 = 0.1$; (C, D) $\Gamma_1 = 1$ and $\Gamma_2 = 0.5$; and (E) $\Gamma_1 = 1$ and $\Gamma_2 = 1$. The total number of atoms $2N$ varies discretely as integers $N \in [5, 150]$. Panels (A) and (C) correspond to the first radiation mode, while (B) and (D) show the fluctuations for the second mode. Panel (E) is identical for both modes.

## 5. Data processing

When one of the spontaneous decay rates $\Gamma_{1,2}$ is set to zero, the two-mode rate equation Eq. (1) reduces to a single-mode rate equation that accurately describes Dicke superradiance. The conventional initial condition for this system is that all $N$ two-level atoms are prepared in their excited (upper) state, with none in the ground state. Under this condition, the system requires time to build up macroscopic coherence among all $N$ atoms. Light emission reaches its peak intensity when this buildup is complete, i.e., when $N/2$ atoms are in their excited states and the remaining $N/2$ atoms are in the ground state. Explicitly, the Dicke superradiance exhibits the following main characteristics: (i) the peak intensity scales as $N^2$; (ii) the maximum peak is delayed by $\tau_D = (E_0 + \log N)/N$; (iii) the time-domain noise fluctuations are given by $\sigma = \pi/[\sqrt{6}(E_0 + \log N)]$; and (iv) the superradiant pulse width scales as $\tau_W \propto 1/N$.

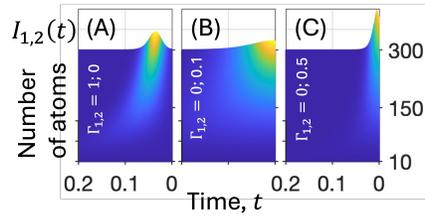

**Figure 4.** Temporal evolution of intensities as functions of the number of atoms for the reduced two-level model derived from the three-level model. Results are shown for the spontaneous decay rates: (A) $\Gamma_1 = 1$ and $\Gamma_2 = 0$; (B) $\Gamma_1 = 0$ and $\Gamma_2 = 0.1$; and (C) $\Gamma_1 = 0$ and $\Gamma_2 = 0.5$. The total number of atoms is $2N$, with $N$ varying discretely as integers $N \in [5, 150]$. Initial condition for (B) and (C): $N$ atoms are in their excited state $|2\rangle$ and the remaining $N$ atoms are in their ground state $|3\rangle$. No transition occurs between $|1\rangle$ and $|3\rangle$. Initial condition for (A): $N$ atoms in their excited state $|1\rangle$ and none is in their ground state $|3\rangle$. No transition occurs between $|2\rangle$ and $|3\rangle$.

To demonstrate these characteristics for the two-level model, we adopt the present model with $2N$ three-level atoms from Eq. (1) but set one of the decay rates to zero. In this reduced two-level model, one of the transitions--either $|2\rangle \rightarrow |3\rangle$ or $|1\rangle \rightarrow |3\rangle$ – is effectively cancelled. Fig. 4 shows the results for the following decay rates: (A) $\Gamma_1 = 1$ and $\Gamma_2 = 0$; (B) $\Gamma_1 = 0$ and $\Gamma_2 = 0.1$; and (C) $\Gamma_1 = 0$ and $\Gamma_2 = 0.5$. For panels (B) and ( C), $N$ atoms are prepared in the upper state $|2\rangle$ and the remaining $N$ atoms are in the lower state $|3\rangle$, with none in the uppermost state $|1\rangle$. This initial condition contrasts with the conventional case in panel (A), where all atoms are initially excited. In this unconventional case, the system is already prepared in a macroscopic coherent state, and thus superradaint emission occurs instantly, as seen in Fig. 4 (B, C).

To better examine the temporal characteristics, the intensities from Fig.4 are normalized and plotted against the theoretical superradiance delay formula $\tau_D = (E_0 + \log N)/N$ in perspective view in Fig. 5. As seen from Fig. 5 (A), the normalized intensity profile for $\Gamma_1 = 1$ and $\Gamma_2 = 0$ reveals the expected linear behavior against the delay formula. Although this formula is derived for a system evolving from all excited atoms (the conventional initial condition), a linear relationship is still visible from Fig. 5 (B, C) even for the unconventional initial condition. This is because the width of superradiant pulse also scales as $1/N$.

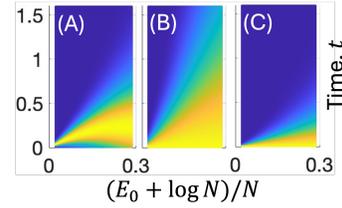

**Figure 5.** Processed data for the reduced two-level model. Normalized intensities from the data presented in Fig. 4 are plotted as functions of time $t$ and delay formula $(E_0 + \log N)/N$. Parameters and panel labels (A-C) are consistent with Fig. 4: (A) $\Gamma_1 = 1$ and $\Gamma_2 = 0$; (B) $\Gamma_1 = 0$ and $\Gamma_2 = 0.1$; and (C) $\Gamma_1 = 0$ and $\Gamma_2 = 0.5$.

These established results for the two-level model are compared with those for the three-level models with various non-zero decay rates. These comparisons provide evidence for superradiant behavior not only in the first mode but also in the second mode of the two-mode superradiance. To facilitate this comparison, we process the data as follows: first, we normalize the intensity data from Fig. 1 and plot them against the delay formula, as was done in Fig. 5. The processed data for the three-level model demonstrate a clear linear relationship, confirming that both modes satisfy one of the main criteria for Dicke superradiance, consistent with the behavior observed in collective two-level systems.

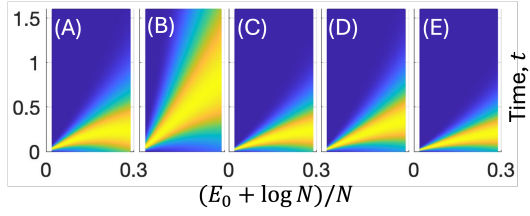

**Figure 6.** The processed data for the three-level model. The normalized intensities for the data presented in Fig. 1 as functions of time $t$ and delay formula $(E_0 + \log N)/N$. Parameters and panel labels (A-E) maintain consistency with Fig. 1 that include: (A, B) $\Gamma_1 = 1$ and $\Gamma_2 = 0.1$; (C, D) $\Gamma_1 = 1$ and $\Gamma_2 = 0.5$; and (E) $\Gamma_1 = \Gamma_2 = 1$.

In the following analysis, we examine whether the two-mode emissions maintain the characteristic $N^2$ scaling. Figure 7 presents the peak intensities extracted from the data in Fig. 1, plotted against $N^2$. The peak intensities for both modes demonstrate linear scaling across the full range of numbers ($N = 10 - 300$) for all three cases: (A) $\Gamma_1 = 1$ and $\Gamma_2 = 0.1$; (B) $\Gamma_1 = 1$ and $\Gamma_2 = 0.5$ and (C) $\Gamma_1 = 1$ and $\Gamma_2 = 1$. These data are also compared to the results extracted from Fig. 4 for the reduced two-level model. In case (A), where the ratio of decay rates is lowest $\Gamma_2/\Gamma_1 = 0.1$, the results for both modes agree well with the corresponding two-level model data, as seen in Fig. 7 (A). Specifically, the first mode (red dashed curve) with $\Gamma_1 = 1$ and $\Gamma_2 = 0$, from Fig. 4 (A), and the second mode (black dashed curve) with $\Gamma_1 = 0$ and $\Gamma_2 = 0.1$, from Fig. 4 (B) are in good agreement with the first (red solid curve) and second (black dash-dotted curve) modes for $\Gamma_1 = 1$ and $\Gamma_2 = 0.1$. As the ratio increases, this agreement no longer holds, as shown in panels (B) and (C) for ratios 0.5 and 1, respectively. It is worth noting that in the Λ-type three-level system, only one of the two modes satisfy the criteria for Dicke superradiance, while the other is diminished [44]. In contrast, for the V-type system, both modes exhibit a clear linear relationship with $N^2$. This confirms that both modes satisfy another key criterion for Dicke superradiance, consistent with the behavior observed in collective two-level systems.

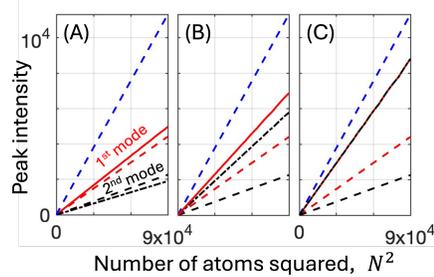

**Figure 7.** Processed data. Peak intensities as function of $N^2$ for the three-level and reduced two-level models. The solid red and black dash-dotted curves represent the first and second modes, respectively, for the three-level model with parameters: (A) $\Gamma_1 = 1$ and $\Gamma_2 = 0.1$ (from Fig. 1 (A); (B) $\Gamma_1 = 1$ and $\Gamma_2 = 0.5$ extracted from Fig. 1 (B); and (C) $\Gamma_1 = 1$ and $\Gamma_2 = 1$ extracted from data in Fig. 1 (C). In all three panels, the dashed curve with the greatest slope (blue) corresponds to the second mode with

$\Gamma_1 = 0$ and $\Gamma_2 = 0.5$, extracted from data in Fig. 4 (C), the dashed curve with the moderate slope (red) corresponds to the first mode with $\Gamma_1 = 1$ and $\Gamma_2 = 0$, extracted from data in Fig. 4 (A), and the dashed curve with the smallest slope (black) corresponds to the second mode with $\Gamma_1 = 0$ and $\Gamma_2 = 0.1$, extracted from data in Fig. 4 (B).

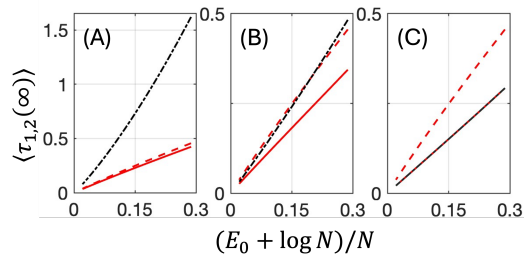

**Figure 8.** Processed data. Asymptotic delay times $\langle \tau_{1,2}(\infty) \rangle$ extracted from Fig. 2 data, plotted against the theoretical superradiance delay $(E_0 + \log N)/N$. The results for the first (solid red curve) and second (dash-dotted black curve) modes: (A) $\Gamma_1 = 1$ and $\Gamma_2 = 0.1$ extracted from data in Fig. 2 (A); (B) $\Gamma_1 = 1$ and $\Gamma_2 = 0.5$ extracted from data in Fig. 2 (B); and (C) $\Gamma_1 = 1$ and $\Gamma_2 = 1$ extracted from data in Fig. 2 (C). In all three panels, the dashed curve (red) corresponds to the first mode with $\Gamma_1 = 1$ and $\Gamma_2 = 0$, extracted from Eq. (6).

The next stage of data processing involves a quantitative analysis in the time domain. We focus on the delay time data from Fig. 2 rather than the intensity profiles of Figs. 1 or 5. As in Ref. [44], the temporal evolution in Fig. 2 reveals two distinct dynamical regimes: (i) an initial rapid transient stage, followed by (ii) asymptotic convergence to steady values in the $t \to \infty$ limit. Figure 8 presents the steady-state delays $\langle \tau_{1,2}(\infty) \rangle$ extracted from Fig. 2. Notably, the data for the first mode with $\Gamma_1 = 1$ and $\Gamma_2 = 0.1$ (see overlapping red solid and red dashed curves, Fig. 7 A) show exact correspondence with the theoretical prediction $\tau_D = (E_0 + \log N)/N$. This agreement demonstrates that in the limit $\Gamma_2/\Gamma_1 \ll 1$, the V-type three-level system effectively reduces to a two-level system, recovering conventional Dicke superradiance behavior. This contrasts with the Λ-type system [44], where the first mode exhibits Dicke superradiance, but the second mode is completely depleted due to the shared excited states. These deformations arise from intermodal competition—the upper state population leaks through both decay channels simultaneously, disrupting the conditions required for pure superradiance in either mode. In contrast, for the present V-type system, intermodal competition is less pronounced. As a result, both modes exhibit Dicke superradiance behavior in the time-domain, as shown by the linearly behaving black dash-dotted curve and red solid curves in all three panels of Fig. 8. For validation, additional results obtained from Eq. (6) with $\Gamma_1 = 1$ and $\Gamma_2 = 0$, are included as red dashed curves in all three panels of Fig. 8.

The subsequent analysis examines the temporal evolution of the two-mode quantum noise fluctuations in the time domain. The processed data are compared with the theoretical prediction $\sigma = \pi/[\sqrt{6}(E_0 + \log N)]$. From the data in Fig. 3, both the asymptotic noise values $\langle\sigma_{1,2}(\infty)\rangle$ and their temporal minima are extracted and plotted against the analytical noise expression in Fig. 9. The numerical results demonstrate consistent with theory across the entire parameter space, including all examined atom numbers ($N = 10$ to 300) and decay rates ($\Gamma_2 = 0.1, 0.5, 1$ and $\Gamma_1 = 1$). Similar to the $\Lambda$-type three-level system, this agreement persists despite the system's inherent dynamical evolution of fluctuations, confirming that the fundamental noise characteristics of Dicke superradiance remain preserved also in V-configuration system. Both the asymptotic values and fluctuation minima (the latter represented by dotted curves with circles in Fig. 9) qualitatively follow the predicted theoretical scaling. This consistent validation of the noise scaling relationship holds true for both radiation modes.

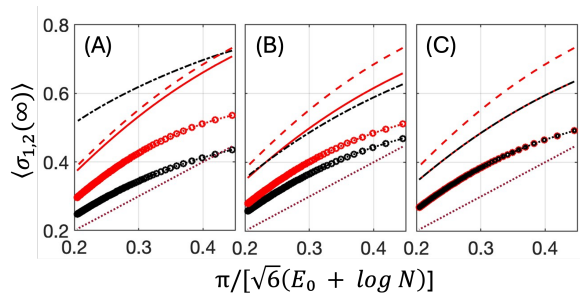

**Figure 9.** Processed data. Asymptotic two-mode quantum noise fluctuations $\langle\sigma_{1,2}(\infty)\rangle$ and their minimum values extracted from Fig. 3, plotted versus the theoretical noise limit $\pi/[\sqrt{6}(E_0 + \log N)]$. The results for the first (solid red curve) and second (dash-dotted black curve) modes for $\langle\sigma_{1,2}(\infty)\rangle$: (A) $\Gamma_1 = 1$ and $\Gamma_2 = 0.1$ extracted from data in Fig. 3 (A); (B) $\Gamma_1 = 1$ and $\Gamma_2 = 0.5$ extracted from data in Fig. 3 (B); and (C) $\Gamma_1 = 1$ and $\Gamma_2 = 1$ extracted from data in Fig. 3 (C). The results for the first (red dotted curve with red circles) and second (red dotted curve with red circles) modes for minima of time-dependent noise fluctuations $\langle\sigma_{1,2}(t)\rangle$. In all three panels, the identical dashed curves (red) correspond to the first mode with $\Gamma_1 = 1$ and $\Gamma_2 = 0$, extracted from Eq. (7), while identical dotted lines for $\pi/[\sqrt{6}(E_0 + \log N)]$ versus $\pi/[\sqrt{6}(E_0 + \log N)]$.

## 5. Superradiant syntheses

For the $\Lambda$-type three level system [44], we demonstrated that competition between the two modes enables mode-selective cooperative emissions, resulting in pure Dicke superradiance in one mode and a deformed emission in the other. By selecting the decay rate ratio, one mode can be selectively enhanced or suppressed. However, this effect is much less pronounced in ladder three-level systems [30] due to the sequential (cascade) nature of their transitions. For instance, with highly imbalanced decay rates such as $\Gamma_1 = 1$ and $\Gamma_2 = 0.01$, a sequential pair of pure Dicke superradiant pulses can be produced in the two modes. It is important to note that in the ladder system, a time-ordered pair of superradiance pulses is produced under conventional initial conditions. This means that the first mode emission is completed within the first mode delay time $\tau_{1D} = (E_0 + \log N)/N$, with total $N$ atoms. The second mode emission then occurs much later, approximately after the second mode superradiant delay time $\tau_{2D} = (\Gamma_1/\Gamma_2)(E_0 + \log N)/N$. The total completion time is longer than sum $\tau_{1D} + \tau_{2D}$. This process is referred to as cascade superradiance [30]. For example, for a total $N$=300 atoms and $\Gamma_1 = 1$, $\Gamma_2 = 0.1$, the delay times are estimated to be $\tau_{1D} \sim 0.02$ and $\tau_{2D} \sim 0.21$ giving sum $\tau_{1D} + \tau_{1D} \sim 0.23$.

In this work, we demonstrate generation of a pair of superradiance pulses under an unconventional initial condition via the V-three level system of $2N$. This means that the first mode emission ($N$ atoms in one upper state) is completed after the first mode delay time $\tau_{1D} = (E_0 + \log N)/N$, while the second mode emission occurs instantly. The total completion time is, therefore, about $\tau_{1D}$, but not $\tau_{1D} + \tau_{2D}$. We refer to this speedup process as a superradiant synthesis. Let us explain this process in detail. First, the processed data for this special case satisfy all the criteria for Dicke superradiance mentioned in this work. Explicitly, (i) the peak intensity is scaled as $N^2$, rather than $N$, see Fig. 7 (A); (ii) peak is delayed by $\tau_D = (E_0 + \log N)/N$, see Fig. 8 (A) also Fig. 6 (A, B); (iii) the time-domain noise fluctuations are given by $\sigma = \pi/[\sqrt{6}(E_0 + \log N)]$, see Fig. 9 (A) ; and (iv) the superradiant pulse width is scaled as $\tau_W \propto 1/N$, see Fig. 6 (A, B). Therefore, the two-level model provides an acceptable framework for interpreting the data from the V-type three-level system.

Figure 10 illustrates this mechanism, plotting the two-mode intensities obtained from Eq. (2) as functions of time and the number of atoms. The surfaces represent the following. Blue surface (A, B): first mode intensity for $\Gamma_1 = 1$ and $\Gamma_2 = 0.1$ under the conventional initial condition. Cyan surface (A, C): second mode intensity for $\Gamma_1 = 1$ and $\Gamma_2 = 0.1$ under the conventional initial condition. Green surface (A, B): first mode intensity for $\Gamma_1 = 1$ and $\Gamma_2 = 0$ under the conventional initial condition. Red surface (A, C): second mode intensity for $\Gamma_1 = 0$ and $\Gamma_1 = 0.1$ under the unconventional initial condition. The result for the unconventional condition (red surface) is plotted at a delay time given by $\alpha(E_0 + \log N)/N$, using a scaling factor of $\alpha = 1/10$ as discussed in [30].

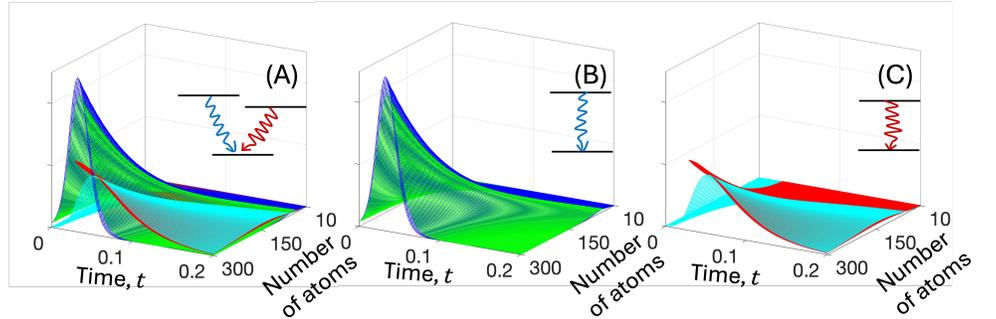

**Figure 10.** A Superradiant synthesis. The intensities given by Eq. (2) are plotted as functions of time and number of atoms. Blue surface (A, B): first mode intensity for $\Gamma_1 = 1$ and $\Gamma_2 = 0.1$ under the conventional initial condition. Cyan surface (A,

C): second mode intensity for $\Gamma_1 = 1$ and $\Gamma_2 = 0.1$ under the conventional initial condition. Green surface (A, B): first mode intensity for $\Gamma_1 = 1$ and $\Gamma_2 = 0$ under the conventional initial condition. Red surface (A, C): second mode intensity for $\Gamma_1 = 0$ and $\Gamma_1 = 0.1$ under the unconventional initial condition. The result for the unconventional condition (red surface) is plotted at a delay time given by $\alpha(E_0 + log\ N)/N$, using a scaling factor of $\alpha = 1/10$ as discussed in [30].

To estimate the delay time, consider a system with $2N = 300$ V-type atoms, with $\Gamma_1 = 1$ and $\Gamma_2 = 0.1$. The system starts with $N = 150$ atoms in the uppermost state $|1\rangle$, another $N = 150$ atoms in the other upper state $|2\rangle$, and no atoms in the lower state $|3\rangle$. Using these parameters, we obtain $\tau_{1D} = (E_0 + log\ N)/N$~0.04. Thus, the first mode emission is completed in about $\tau_{1D}$~0.04, and the second mode emission occurs virtually instantly. The overall waiting time is therefore $\tau_{1D}$~0.04, which is nearly an order of magnitude faster than the cascade process time of $\tau_{1D} + \tau_{2D}$~ 0.23. In this synthesis, one mode of superradiance occurs in the transition $|1\rangle \rightarrow |3\rangle$ driven by maximum macroscopic coherence among 150 atoms. This is immediately followed by another mode of superradiance in the transition $|2\rangle \rightarrow |3\rangle$ driven by maximum macroscopic coherence among all 300 atoms. The atomic configuration consists of two distinct upper states and a shared lower state, where one-photon transitions are permitted only between the common lower state and each upper state, but are forbidden between the two upper states. This means that the two sub-ensembles, each with $N$ atoms, are initially dynamically decoupled but later synthesize into a single ensemble of $2N$ two-level atoms. This is the main reason we use the term 'synthesis' in this context.

     In conclusion, we have presented an extension of the Dicke superradaince model from a two-level model to a V-type three-level model. We investigated the situation in which $N$ atoms are in one excited state, another $N$ atoms are in the second excited state, and no atoms occupy the common ground state. Under this initial condition, we obtained numerical data and analyzed it against the principal criteria for superradaince. When the two spontaneous decay rates differ by more than an order of magnitude, the system—initially composed of dynamically decoupled sub-ensembles—emits a well-timed, successive pair of superradiant pulses. This behavior effectively synthesizes the two separate sub-ensembles of $N$ atoms each into a single macroscopic ensemble of $2N$ atoms. We term this novel phenomenon superradiant synthesis. This finding offers new insights for practical quantum state engineering, particularly by enabling the synthesis of macroscopic quantum subsystems.

funding acquisition, G.O.A. and T.B. All authors have read and agreed to the published version of the manuscript.

**Funding:** T.B.'s research was partially funded by the National University of Mongolia, grant number P2022-4375.

**Data Availability Statement:** Partial data are available upon request.

**Conflicts of Interest:** The authors declare no conflicts of interest.